# On the Equality of Complexity Classes *P* and *NP*: Linear Programming Formulation of the Quadratic Assignment Problem

Moustapha DIABY, OPIM Dept., University of Connecticut, Storrs, CT 06268, USA

*Abstract* - In this paper, we present a first linear programming formulation of the *Quadratic Assignment Problem* (QAP). The proposed linear program is a network flow-based model with $O(n^9)$ variables and $O(n^7)$ constraints, where n is the number of assignments. Hence, it provides for the solution of the QAP in polynomial-time and represents therefore, a proof of the equality of the computational complexity classes *P* and *NP*. Computational testing and results are discussed.

*Index Terms* - Linear Programming; Combinatorial Optimization, Quadratic Assignment Problem, Facility Location, Facility Layout, Computational Complexity, Integer Programming.

## I. INTRODUCTION

The Quadratic Assignment Problem (QAP) is the problem of making exclusive assignments of *n* indivisible entities to *n* other indivisible entities in such a way that a total quadratic interaction cost is minimized. The problem can be interpreted from a wide variety of perspectives. The perspective we adopt in this paper is that of the generic facilities location/layout context, as in the seminal work of Koopmans and Beckmann [1]. Specifically, there are n facilities (or departments) to be located at n possible sites (or locations). The volume of traffic going from facility i to facility j is denoted $f_{ij}$. The travel distance from site r to site s is denoted $d_{rs}$. A quadratic "material handling" cost of $h_{irjs} = (f_{ij}d_{rs} + f_{ji}d_{sr})$ is incurred if facilities i and j are assigned to sites r and s, respectively. In addition, there is a fixed cost (an "operating cost"), $o_{ir}$, associated with operating facility i at site r. It is assumed (without loss of generality) that the units for "distance", "volume of traffic", and "operating cost" have been chosen so that the $h_{irjs}$'s and $o_{ir}$'s are commensurable. The problem is that of finding a perfect matching of the facilities and sites so that the total material handling and facilities operating costs is minimized.

Let $F = \{1, 2, ..., |F|\}$ and $S = \{1, 2, ..., |S|\}$ be the sets of facilities and sites, respectively. Without loss of generality, assume $|F| = |S| = n$. For $i \in F$ and $r \in S$, let $w_{ir}$ be a 0/1 binary variable that indicates whether facility i is assigned to (or located at) site r ($w_{ir} = 1$), or not ($w_{ir} = 0$). Then, a classical formulation of the QAP is as follows:

**Problem QAP**:

Minimize
$$Z_{QAP}(w) = \sum_{i \in F} \sum_{j \in F} \sum_{r \in S} \sum_{s \in S} h_{irjs} w_{ir} w_{js} + \sum_{i \in F} \sum_{r \in S} o_{ir} w_{ir} \quad (1.1)$$
Subject to:
$$\sum_{i \in F} w_{ir} = 1 \quad r \in S \quad (1.2)$$
$$\sum_{r \in S} w_{ir} = 1 \quad i \in F \quad (1.3)$$
$$w_{ir} \in \{0, 1\} \quad i \in F; \quad r \in S \quad (1.4)$$

Problem QAP was shown to be NP-Hard as far back as the 1970's (see [2]). Moreover, it has been known for some time that the Traveling Salesman Problem (see [3]) and other *NP-Complete* combinatorial optimization problems (see [4], [5], or [6]) can be modeled as special cases of the problem. Hence, the thrust of research on the problem has been towards the development of heuristic procedures and "tight" lower bounds (see [7], [8], and [9] for reviews).

In this paper, we present a first linear programming formulation of the *Quadratic Assignment Problem* (QAP). The proposed linear program is a network flow-based model with $O(n^9)$ variables and $O(n^7)$ constraints, where n is the number of assignments. Hence, it provides for the solution of the QAP in polynomial-time and represents therefore, a proof of the equality of the computational complexity classes *P* and *NP*. Computational testing and results are discussed.

The plan of the paper is as follows. We develop the proposed linear programming formulation in section 2. Computational testing and results are discussed in section 3. Conclusions are discussed in section 4.

## II. DEVELOPMENT OF THE FORMULATION

In this section, we first develop a network flow-based Integer Linear Programming (ILP) formulation of the QAP. Then, we discuss the development of our Linear Programming (LP) formulation.

### 2.1 Integer Linear Programming Model

The basic idea of our modeling is to express the polytope associated with *Problem QAP* in terms of higher-dimensional variables in such a way that the quadratic cost function of *Problem QAP* is correctly captured using a linear function. Note that this polytope (i.e., the polytope associated with *Problem QAP*) is the standard assignment polytope (see [10], or [11]). To reformulate this polytope we use the framework of the multipartite graph $G = (V, A)$ illustrated in Figure 1.1, where the nodes in V correspond to (facility, site) pairs, and the arcs in A correspond to binary variables $x_{irj} \equiv w_{ir} w_{j,r+1}$ ($(i, j) \in F^2; r \in S\setminus\{n\}$). Clearly, there is a one-to-one correspondence between feasible solutions to *Problem QAP* (i.e., perfect matchings of the facilities and sites) and paths in Graph G that simultaneously span the set of facilities and the set of sites of the graph, respectively. Hence, we refer to such paths as "*perfect bipartite matching (p.b.m.) paths*." Our reformulation approach consists of developing constraints that "force" flow propagation in Graph G to occur along *p.b.m. paths* of the graph only. In order to simplify the presentation, we refer to the set of all the nodes of the graph that have a given facility index in common as a "level" of the graph and to the set of all the nodes of the graph that have a given site index in common as a "stage" of the graph.

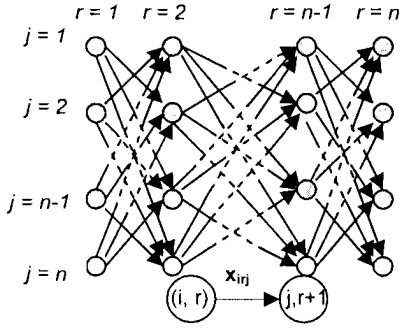

Figure 1.1: Network Sub-Structure of Problem ILP

Our proposed overall model is a more general form of that developed in [12]. Define $R \equiv S \setminus \{n\}$. For $(i, j, k, t, u, v) \in F^6$, $(p, r, s) \in R^3$ such that $p < r < s$, let $z_{upvirjkst}$ be a 0/1 binary variable that takes on the value "1" if and only if the flow on arc $(u, p, v)$ of Graph G subsequently flows on arcs $(i, r, j)$ and $(k, s, t)$, respectively. Similarly, for $(i, j, k, t) \in F^4$, $(r, s) \in R^2$ such that $s > r$, let $y_{irjkst}$ be a binary variable that indicates whether the flow on arc $(i, r, j)$ subsequently flows on arc $(k, s, t)$ ($y_{irjkst} = 1$) or not ($y_{irjkst} = 0$). Finally, denote by $y_{irjirj}$ the binary variable that indicates whether there is flow on arc $(i, r, j)$ of Graph G or not. Given an instance, $(y, z)$, of these decision variables, we use the term "flow layer" to refer to the sub-graph of G induced by the arc $(i, r, j)$ corresponding to a given positive component, $y_{irjirj}$, of $(y)$ and the corresponding arcs $(k, s, t)$ $(s \in R, s > r)$ such that $y_{irjkst} > 0$. Hence, the flow on arc $(i, r, j)$ also flows on arc $(k, s, t)$ (for a given $s > r$) iff arc $(k, s, t)$ belongs to the flow layer originating from arc $(i, r, j)$. Also, we say that flow on a given arc $(i, r, j)$ of Graph G "visits" a given level of the graph, say level $t$, if:

$$\sum_{s \in R; s \leq r-1} \sum_{k \in (F \setminus \{i,j,t\})} y_{tskirj} + \sum_{s \in R; s \geq r+1} \sum_{k \in (F \setminus \{i,j,t\})} y_{irjkst} > 0 \quad (2.1)$$

Logical constraints of our model are that: 1) flow must be conserved; 2) flow must be connected; and, 3) flow layers must be consistent with one another. By "consistency" of the flow layers, we are referring to the requirement that any flow layer originating from a given arc $(i, r, j)$ with $r \geq 2$ must be a subgraph of one or more flow layers originating from a set of arcs at any other given stage preceding $r$. More specifically, consider the arc $(i, r, j)$ corresponding to a given positive component of $(y)$, $y_{irjirj} > 0$. For $s < r$ $(s \in R)$, define $F_s(i, r, j) \equiv \{(k, t) \in F^2 \mid y_{kstirj} > 0\}$. Then, by "consistency of flow layers" we are referring to the condition that the flow layer originating from arc $(i, r, j)$ must be a sub-graph of the union of the flow layers originating from the arcs comprising each of the $F_s(i, r, j)$'s, respectively. In addition to the logical constraints, the bipartite matching constraints 1.2 and 1.3 of Problem QAP must be respectively enforced. These ideas are developed in the following.

1) *Flow Conservations:*

All flows through Graph G must be initiated at *stage 1*; Also, for $(i, j) \in F^2$, $r \in (R \setminus \{1\})$, the flow on arc $(i, r, j)$ must be equal to the sum of the flows from *stage 1* that subsequently flow on $(i, r, j)$.

$$\sum_{i \in F} \sum_{j \in F} y_{i,1,j,i,1,j} = 1 \quad (2.2)$$

$$y_{irjirj} - \sum_{u \in F} \sum_{v \in F} y_{u,1,virj} = 0; \quad i, j \in F; \quad r \in R, \quad r \geq 2 \quad (2.3)$$

2) *Flow Connectivities:*

All flows must propagate through the graph, from *stage* 1 on to *stage* n, in a connected manner; Each *flow layer* must be a connected graph and must conserve flow.

$$\sum_{i \in F} y_{i,r-1,j,i,r-1,j} - \sum_{i \in F} y_{jrjri} = 0; \quad r \in R, r \geq 2; j \in F \quad (2.4)$$

$$\sum_{k \in F} y_{irjkst} - \sum_{k \in F} y_{irjt,s+1,k} = 0; \quad i, j, t \in F;$$

$$r, s \in R, \quad r \leq n-2, r \leq s \leq n-2 \quad (2.5)$$

3) *Consistency of "Flow Layers:"*

For $r, s \in R, r < s$, flow on $(i, r, j)$ subsequently flows onto $(k, s, t)$ iff for each $p < r$ $(p \in R)$ there exists at least one pair $(u, v) \in F^2$ such that flow from $(u, p, v)$ propagates onto $(k, s, t)$ via $(i, r, j)$. This results in the following three types of constraints:

*i) Layering Constraints A*

$$y_{upvirj} - \sum_{k \in F} \sum_{t \in F} z_{upvirjkst} = 0; \quad u, v, i, j \in F;$$

$$p, r, s \in R, 2 \leq r \leq n-2; \quad p \leq r-1; \quad s \geq r+1 \quad (2.6)$$

*ii) Layering Constraints B*

$$y_{irjkst} - \sum_{u \in F} \sum_{v \in F} z_{upvirjkst} = 0; \quad u, v, i, j \in F;$$

$$p, r, s \in R, 2 \leq r \leq n-2; p \leq r-1; \quad s \geq r+1 \quad (2.7)$$

*iii) Layering Constraints C*

$$y_{upvkst} - \sum_{i \in F} \sum_{j \in F} z_{upvirjkst} = 0; \quad u, v, i, j \in F;$$

$$p, r, s \in R, 2 \leq r \leq n-2; p \leq r-1; \quad s \geq r+1 \quad (2.8)$$

4) *"Visit" Requirements:*

Flow within any *layer* of Graph G must *visit* every *level* of the graph.

$$y_{u,1,vu,1,v} - \sum_{s \in R; s \geq 2} \sum_{k \in F} y_{u,1,vkst} = 0;$$

$$u, v \in M; \quad t \in F \setminus \{i, j\} \quad (2.9)$$

$$y_{u,1,virj} - \sum_{s \in R; s \leq r-1} \sum_{k \in F} z_{u,1,vtskirj} - \sum_{s \in R; s \geq r+1} \sum_{k \in F} z_{u,1,virjkst} = 0;$$

$$r \in R \setminus \{1\}; \quad u, v, i, j \in F; \quad t \in (F \setminus \{u, v, i, j\}) \quad (2.10)$$

5) *"Visit" Restrictions:*

Flow must be connected with respect to the *stages* of Graph G; There can be no flow between nodes belonging to the same *level* of the graph; No *level* of the graph can be visited at more than one *stage*, and vice versa.

$$\sum_{s \in R; s < r} \sum_{k \in F} \sum_{t \in F} y_{irjkst} + \sum_{(k,t) \in F^2 \mid (k,t) \neq (i,j)} y_{irjkrt} +$$

$$\sum_{k \in (F \setminus \{j\})} \sum_{t \in F} y_{irjk,r+1,t} + \sum_{s \in R; s \geq r+1} \sum_{k \in F} y_{irjksi} + \sum_{s \in R; s \geq r+1} \sum_{k \in F} y_{irjisk} +$$

$$\sum_{s \in R; s \geq r+1} \sum_{k \in F} y_{irjksj} + \sum_{s \in R; s \geq r+2} \sum_{k \in F} y_{irjjsk} + \sum_{s \in R} \sum_{k \in F} \sum_{t \in F} y_{iriksrt}$$

$$+ \sum_{s \in R} \sum_{k \in F} \sum_{t \in F} y_{kstjrj} = 0; \quad i, j \in F; \quad r \in R \quad (2.11)$$

Note that constraints 1.2 of Problem QAP are enforced through the combination of the "Flow Connectivities" requirements and the "Visit Restrictions" constraints, and that constraints 1.3 are enforced through the *"Visit Requirements"* constraints.

Let $c_{irj}$ $((i, j) \in F^2; r \in R)$ be defined as:

$$c_{irj} \equiv \begin{cases} o_{ir} + f_{ij}d_{r,r+1} + f_{ji}d_{r+1,r} \\ \quad \text{for } r \in R, \ r \leq n-2; i \in F, j \in F \setminus \{i\} \\ o_{ir} + o_{j,r+1} + f_{ij}d_{r,r+1} + f_{ji}d_{r+1,r} \\ \quad \text{for } r = n-1; i \in F, j \in F \setminus \{i\} \\ \infty, \quad \text{otherwise} \end{cases} \quad (2.12)$$

Then, our integer linear programming model can be stated as follows:

**Problem ILP:**
Minimize

$$Z_{IP}(y, z) = \sum_{i \in F} \sum_{r \in (R \setminus (n-1))} \sum_{t \in F} \sum_{s \in R; s > r} \sum_{j \in F} \sum_{k \in F} h_{irt,s+1} y_{irjkst} + \sum_{i \in F} \sum_{r \in R} \sum_{j \in F} c_{irj} y_{irjirj} \quad (2.13)$$

Subject to:

Constraints 2.2 – 2.11

$y_{irjkst}, z_{upvirjkst} \in \{0, 1\}, \quad j, k, t, u, v \in F; \ p, r, s \in R \quad (2.14)$

We formally establish the equivalence between *Problem ILP* and *Problem QAP* in the following proposition, the proof of which is given in [13].

**Proposition 1**
*Problem ILP* and *Problem QAP* are equivalent.

Hence, each feasible solution to *Problem ILP* corresponds to a perfect bipartite matching solution of *Problem QAP*, and therefore, to a *p.b.m.* path in Graph G, and conversely. Let $\varphi(\ell) = \{\ell_1, \ell_2, \cdots, \ell_{n-1}, \ell_n\}$ denote the ordered set of facility indices corresponding to a given perfect matching, $\ell$, of the facilities and sites (i.e., with $\ell_t$ as the index of the facility assigned to site t according to $\ell$). In the remainder of this paper, we will use the term *"feasible solution corresponding to ((Given) Perfect Matching) $\ell$"* to refer to the vector $(y(\varphi(\ell)), z(\varphi(\ell)))$ obtained as follows:

$$(y(\varphi(\ell)))_{arbcsd} = \begin{cases} 1 & \text{for } r, s \in R, \ s \geq r; \\ & (a,b,c,d) = (\ell_r, \ell_{r+1}, \ell_s, \ell_{s+1}), \\ 0 & \text{otherwise} \end{cases} \quad (2.15)$$

$$(z(\varphi(\ell)))_{apbcrdesf} = \begin{cases} 1 & \text{for } p, r, s \in R, \ p < r < s; \\ & (a, b, c, d, e, f) = \\ & (\ell_p, \ell_{p+1}, \ell_r, \ell_{r+1}, \ell_s, \ell_{s+1}); \\ 0 & \text{otherwise} \end{cases} \quad (2.16)$$

The following proposition gives some further characterization of the feasible set of *Problem ILP* (The proof is given in [13]).

**Proposition 2**
The following constraints are valid for *Problem ILP*:

i) $y_{irjirj} - \sum_{k \in F} \sum_{t \in F} y_{kstirj} = 0 \quad i, j \in F; \ r, s \in R,$

$r \geq 2, \ s \leq r-1 \quad (2.17)$

ii) $y_{irjirj} - \sum_{k \in F} \sum_{t \in F} y_{irjkst} = 0 \quad i, j \in F; \ r, s \in R,$

$r \leq n-2, \ s \geq r+1 \quad (2.18)$

iii) $\sum_{u \in F} \sum_{v \in F} z_{ugvirjkst} - \sum_{u \in F} \sum_{v \in F} z_{irjupvkst} = 0 \quad i, j, k, t \in F;$

$g, p, r, s \in R; \ g < r < p < s \quad (2.19)$

iv) $\sum_{u \in F} \sum_{v \in F} z_{ugvirjkst} - \sum_{u \in F} \sum_{v \in F} z_{irjkstuqv} = 0 \quad i, j, k, t \in F;$

$g, q, r, s \in R; \ g < r < s < q \quad (2.20)$

v) $\sum_{u \in F} \sum_{v \in F} z_{irjupvkst} - \sum_{u \in F} \sum_{v \in F} z_{irjkstuqv} = 0 \quad i, j, k, t \in F;$

$p, q, r, s \in R; \ r < p < s < q \quad (2.21)$

## 2.2 Linear Programming Model

Our overall linear programming model consists of the linear programming (LP) relaxation of *Problem ILP*. This problem can be stated as follows:

**Problem $\overline{ILP}$:**

Minimize

$$Z_{LP}(y, z) = \sum_{i \in F} \sum_{r \in R \setminus (n-1)} \sum_{t \in F} \sum_{s \in R, s > r} \sum_{j \in F} \sum_{k \in F} h_{irt,s+1} y_{irjkst} + \sum_{i \in F} \sum_{r \in R} \sum_{j \in F} c_{irj} y_{irjirj} \quad (2.25)$$

Subject to:

Constraints 2.2 – 2.11

$0 \leq y_{irjkst}, z_{upvirjkst} \leq 1; \ u, v, i, j, k, t \in F; \ p, r, s \in R \quad (2.26)$

For a feasible solution $(y, z) = (y_{irjkst}, z_{upvirjkst})$ to *Problem $\overline{ILP}$*, let $G(y, z) = (V(y, z), A(y, z))$ be the sub-graph of G induced by the arcs of Graph G corresponding to the positive components of $(y)$. For $r \in R$, define $X_r(y, z) = \{(i, j) \in F^2 \mid \{(i, r, j) \in A(y, z)\}$. Denote the arc corresponding to the $v^{th}$ component of $X_r(y, z)$ $(v \in \{1, 2, \cdots, \chi_r(y, z)\}; 1 \leq \chi_r(y, z) \leq n(n-1))$ as $a_{r,v}(y, z) = (i_{r,v}, r, j_{r,v})$. Then $X_r(y, z)$ can be alternatively represented as $X_r(y, z) = \{(i_{r,v}, r, j_{r,v}); v \in N_r(y, z)\}$, where $N_r(y, z) = \{1, 2, \cdots, \chi_r(y, z)\}$ is the index set for the arcs of Graph $G(y, z)$ originating at stage r.

We have the following.

**Proposition 3**
Let $(y, z) = (y_{isjkrt}, z_{upvirjkrt})$ be a feasible solution to *Problem $\overline{ILP}$*. For $(r, s) \in R^2$, $s > r; \tau \in N_r(y, z)$; and $\mu \in N_s(y, z)$; if $y_{i_{r,\tau}, r, j_{r,\tau}, i_{s,\mu}, s, j_{s,\mu}} > 0$, then, there must exist at least one sequence of arcs of $G(y, z)$,

$P_{rsr}(y, z) \equiv \{a_{r,\tau}, a_{r+1,v_{r+1,\ell}}, \cdots, a_{s-1,v_{s-1,\ell}}, a_{s,\mu} \mid v_{q,\ell} \in N_q(y, z);$

$q \in R, \ r+1 \leq q \leq s-1 \},$

such that:

i) $i_{q+1, v_{q+1,\ell}} = j_{q, v_{q,\ell}} \quad \text{for } q \in R; \ r \leq q \leq s$

ii) $y_{i_{p,v_{p,\ell}}, p, j_{p,v_{p,\ell}}, i_{q,v_{q,\ell}}, q, j_{q,v_{q,\ell}}} > 0 \quad \text{for } p, q \in R;$

$r \leq p \leq s-1; \ p+1 \leq q \leq s$

iii) $i_{p, v_{p,\ell}} \neq i_{q, v_{q,\ell}} \quad \text{for } (p, q) \in (S \cap [r, s+1])^2; \ p \neq q$

where $v_{r,\ell} = \tau$, and $v_{s,\ell} = \mu$.

*Proof:*
i) Condition i) follows from the flow conservation and flow connectivity requirements stipulated by constraints 2.2 – 2.5;
ii) Condition ii) follows from condition i), constraints 2.17 – 2.21, and the *visit* requirements constraints 2.9 – 2.10;
iii) Condition iii) follows from the combination of condition ii) and the *visit* restrictions constraints 2.11.

Q.E.D.

We say that a set of arcs of $G(y, z)$, $L_t(y, z) = \{a_{1,v_{1,t}}, \cdots, a_{n-1,v_{n-1,t}} \mid v_{r,t} \in N_r(y, z), r \in R\}$, is a "path in $(y, z)$" if $i_{r+1,v_{r+1,t}} = j_{r,v_{r,t}}$ for all $r \in R$. Hence, a *path in (y, z)* can be alternatively represented as an ordered set of facility indices, $L_t(y, z) = \{i_{1,v_{1,t}}, i_{2,v_{2,t}}, \cdots, i_{n,v_{n,t}} \mid v_{r,t} \in N_r(y, z), r \in R; \text{ and } i_{n,v_{n,t}} = j_{n-1,v_{n-1,t}}\}$. We will henceforth use this alternative representation for convenience. We refer to a given *path in (y, z)*, $L_t(y, z)$, as "layered" if it satisfies conditions i)-iii) of Proposition 3 above.

To a *path in (y, z)*, $L_t(y, z)$, we attach a "flow value" $\lambda_{i_{1,v_{1,t}}, i_{2,v_{2,t}}, t}(y, z)$ defined as:

$$\lambda_{i_{1,v_{1,t}}, i_{2,v_{2,t}}, t}(y, z) \equiv$$

$$\min_{(p,q) \in (R\setminus\{1\})^2 \mid q>p} \{z_{i_{1,v_{1,t}}, 1, i_{2,v_{2,t}}, i_{p,v_{p,t}}, p, i_{p+1,v_{p+1,t}}, i_{q,v_{q,t}}, q, i_{q+1,v_{q+1,t}}}\} \quad (2.27)$$

A set of *paths in (y, z)*, $\Gamma = \{P_1, P_2, \ldots, P_m\}$ with associated set of arc sets in $G$, $\{a_1, a_2, \ldots, a_m\}$ (where $a_k = \{a_{r,v_{r,k}}; (r, v_{r,k}) \in (R, N_r(y, z))\}$, for $k = 1, \ldots, m$), is said to "cover" $(y, z)$ if $\bigcup_{1 \leq k \leq m}(a_k) = A(y, z)$. Moreover, if $\Gamma$ covers $(y, z)$ with $y_{irjirj} = \sum_{k \mid (i,r,j) \in a_k} \lambda_{i_{1,v_{1,k}}, i_{2,v_{2,k}}, k}(y, z)$ for all (i, r, j) $\in$ A(y, z), then we say that (y, z) "consists of" $\Gamma$. Note that $\lambda_{i_{1,v_{1,t}}, i_{2,v_{2,t}}, t}(y, z) > 0$ iff Path $L_t(y, z)$ is *layered* as described above, that each *layered path in (y, z)* is a *p.b.m. path* of Graph G, and that the *feasible solution corresponding to a given p.b.m. path* of Graph G is a *layered path in (y, z)*.

We will establish the equivalence between *Problem* $\overline{ILP}$ and *Problem QAP* in the remainder of this section.

### Proposition 4
Let $(y, z) = (y_{irjkst}, z_{upvirjkst})$ be a feasible solution to *Problem* $\overline{ILP}$. Then, there exists a set, $\Pi(y, z)$, of perfect matchings of the facilities and sites, such that $(y, z)$ is a convex combination of *feasible solutions corresponding to* the matchings in $\Pi(y, z)$.

*Proof:*
Constraints 2.3 combined with Proposition 3 imply that there exists a set of *layered paths in (y, z)* that *covers (y, z)*. It follows from the correspondence of a given *layered path in (y, z)* to a unique perfect matching of the facilities and sites, and the fact that a given perfect matching of facilities and sites cannot be represented as a convex combination of other perfect matchings of facilities and sites, that (y, z) must *consist* of such a set of *paths in (y, z)*. The proposition follows directly from this.

Q.E.D.

### Proposition 5
The following statements are true of basic feasible solutions (BFS) of *Problem* $\overline{ILP}$ and perfect matchings of the facilities and sites:
1) Every BFS of *Problem* $\overline{ILP}$ corresponds to a perfect matching of the facilities and sites;
2) Every perfect matching of the facilities and sites corresponds to a BFS of *Problem* $\overline{ILP}$;
3) The mapping of BFS's of *Problem* $\overline{ILP}$ onto the set of perfect matchings of the facilities and sites is surjective.

*Proof:*
1) Correspondence of a BFS of *Problem* $\overline{ILP}$ to a perfect matching of the facilities and sites follows from the fact that every perfect matching of the facilities and sites corresponds to a feasible solution to *Problem* $\overline{ILP}$ (Proposition 1), the fact that every feasible solution to *Problem* $\overline{ILP}$ correspond to a convex combination of perfect matching of the facilities and sites (Proposition 4), and the fact that a BFS cannot be a convex combination of other of other feasible solutions;
2) Correspondence of a perfect matching of the facilities and sites to a BFS of *Problem* $\overline{ILP}$ follows from Proposition 1, Proposition 4, and the fact that a given perfect matching of the facilities and sites cannot be represented as a convex combination of other perfect matching of the facilities and sites;
3) The surjective nature of the "BFS's-to-perfect matching of the facilities and sites" mapping follows from the primal degeneracy of *Problem* $\overline{ILP}$.

Q.E.D.

### Corollary 1
*Problem* $\overline{ILP}$ and *Problem ILP* (and therefore, *Problem QAP*) are equivalent.

*Proof:*
The proof follows directly from Proposition 5.

Q.E.D.

### Corollary 2
Computational complexity classes *P* and *NP* are equal.

*Proof:*
First, note that *Problem* $\overline{ILP}$ has $O(n^9)$ variables and $O(n^7)$ constraints. Hence, it can be explicitly stated in polynomial time. The proposition follows directly from this, the NP-Completeness of the QAP decision problem (see [2], and [4]), Corollary 1, and the fact that an explicitly-stated instance of *Problem* $\overline{ILP}$ can be solved in polynomial-time (see [14], and [15]).

Q.E.D.

## III. NUMERICAL IMPLEMENTATION

Because of the very-large-scale nature of *Problem* $\overline{ILP}$, we implemented a streamlined version of it where constraints 2.11 and the variables they restrict to zero were not explicitly considered, and constraints 2.26 were re-written as simple non-negativity constraints (since the upper bounds on the

$y_{irjkst}$ and $z_{upvirjkst}$ variables in those constraints are redundant).

In order to get some idea about the computational performance of our proposed model, we solved 10 randomly-generated 6-facility problems. For each of these problems, the inter-facility traffic volumes were assumed to be uniform random numbers between 10 and 250, and the inter-site distances were assumed to be uniform random numbers between 1 and 30. The facility operating costs were assumed to be zero in five of the problems, and assumed to be random deviates on [0, 5000] for the remainder five problems. In addition to the randomly-generated problems, we also solved one problem where all the inter-site distances were set equal to 10, all the inter-facility traffic volumes were set equal to 50, and all the facility operating costs were set to zero. This additional problem is labeled "QAPn6x."

The computational results are summarized in Table 3.1. (Furher details are provided in [13].) We applied the simplex procedure implementation of the *OSL Optimization Package* (IBM) to solve the dual form of each of the problems. The average computational time (excluding *Problem QAPn6x*) was 16.0814 seconds and 6.7626 seconds of Toshiba Satellite A65-1362, 2.53 GHz Celeron D, Notebook Computer time for the problems without facility operating costs and the problems with facility operating costs, respectively. The corresponding averages of the numbers of iterations were 4,357.8 and 2,705.8, respectively.

We also solved the primal form of each of the test problems. Computational times for the primal form were significantly greater than for the dual LP form in general. However, the primal LP form appeared to hold some promise with respect to future developments because of the relatively small number (specifically, 2, on average) of perfect matchings of the facilities and sites that are examined. Overall, our experimentation with the primal forms provided the empirical validation of our theoretical developments in section 2 of this paper that we were seeking (see Proposition 5, in particular).

## IV. CONCLUSIONS

We have developed a first polynomial-sized linear programming model of the QAP. From a theoretical perspective, the proposed model provides an affirmative resolution to the very long-standing, central, and very far-reaching issue in Operations Research and Mathematics in general, of the equality of computational complexity classes *P* and *NP*. With respect to practice, our proposed model and modeling approach appear to hold some good promises because of the somewhat "friendly," network-based mathematical programming sub-structure of the model, the special ("perfect matching") structure of the basic feasible solutions of the model, and the relatively small number of perfect matchings that are examined when the primal LP form of the model is used.

| Problem Name[1] | Primal Form Number of p.b.m solutions[2] | Dual Form Number of Iterations | Dual Form CPU Seconds[3] | Problem Value |
|---|---|---|---|---|
| QAPn61 | 2 | 3,730 | 11.875 | 60,660 |
| QAPn62 | 3 | 5,036 | 20.563 | 59,924 |
| QAPn63 | 2 | 4,304 | 15.985 | 48,542 |
| QAPn64 | 2 | 4,458 | 16.906 | 44,752 |
| QAPn65 | 1 | 4,261 | 15.078 | 40,525 |
| QAPn6x | 1 | 11,712 | 138.708 | 15,000 |
| **Average**[b] | **2.0** | **4,357.8** | **16.0814** | ---- |
| QAPo61 | 2 | 2,930 | 7.641 | 55,177 |
| QAPo62 | 2 | 2,379 | 5.328 | 51,087 |
| QAPo63 | 1 | 3,293 | 9.516 | 72,720 |
| QAPo64 | 2 | 2,600 | 6.031 | 57,218 |
| QAPo65 | 2 | 2,327 | 5.297 | 53,586 |
| **Average** | **1.8** | **2,705.8** | **6.7626** | ---- |

1: ([a]): "QAPn··": ⇒ operating costs are zero;
   "QAPo··": ⇒ operating costs are positive;
   ([b]): excludes Problem QAPn6x

2: "p.b.m". = "perfect bipartite matching"

3: Total CPU time (Toshiba Satellite A65-S1362 Notebook; 2.53 GHz Celeron D Processor)

*Table 3.1*: Summary of the Computational Results